\documentclass{scrartcl}
\usepackage{amsmath}
\usepackage{amssymb}
\usepackage[round]{natbib}
\usepackage{graphicx}
\usepackage{algorithm,algpseudocode}
\usepackage[colorlinks=true,allcolors=blue]{hyperref}
\usepackage{cleveref}
\usepackage{booktabs}
\usepackage{tabularx}
\usepackage{verbatim}
\usepackage{dcolumn}
\usepackage{orcidlink}
\usepackage{authblk}

\newcommand{\set}[1]{\left\{#1\right\}}
\newcommand{\parenth}[1]{\left(#1\right)}
\newcommand{\bracket}[1]{\left[#1\right]}
\newcommand{\abs}[1]{\left\vert#1\right\vert}
\newcommand{\colalign}[2]{\multicolumn{1}{#1}{#2}}
\newcommand{\email}[1]{\href{mailto:#1}{#1}}

\begin{document}

\title{Bayesian Inference for Sexual Contact Networks Using Longitudinal Survey Data}
\date{}
\author{Till Hoffmann\thanks{\email{thoffmann@hsph.harvard.edu}} \orcidlink{0000-0003-4403-0722} }
\author{Jukka-Pekka Onnela\thanks{\email{onnela@hsph.harvard.edu}} \orcidlink{0000-0001-6613-8668}}
\affil{\normalsize Department of Biostatistics, Harvard T.H.\ Chan School of Public Health}
\maketitle

\begin{abstract}
Characterizing sexual contact networks is essential for understanding sexually transmitted infections, but principled parameter inference for mechanistic network models remains challenging. We develop a discrete-time simulation framework that enables parameter estimation using approximate Bayesian computation. The interpretable model incorporates relationship formation, dissolution, concurrency, casual contacts, and population turnover.
Applying our framework to survey data from 403 men who have sex with men in Stockholm, we provide principled uncertainty quantification for key network dynamics. Our analysis estimates the timescale for seeking a new steady relationship at 25 weeks and for relationship dissolution at 42 weeks. Casual contacts occur more frequently for single individuals (every 1.8 weeks) than for partnered individuals (every 4.5 weeks). However, while cross-sectional data constrains these parameters, migration rates remain poorly identified.
We demonstrate that simple longitudinal data can resolve this issue. Tracking participant retention between survey waves directly informs migration rates, though survey dropout is a potential confounder. Furthermore, simple binary survey questions can outperform complex timeline follow-back methods for estimating contact frequencies. This framework provides a foundation for uncertainty quantification in network epidemiology and offers practical strategies to improve inference from surveys, the primary data source for studying sexual behavior.
\end{abstract}

\section{Introduction}

Characterizing contact networks is essential for understanding and containing infectious disease spread. This is particularly important for sexually transmitted infections (STIs) because structured contact patterns affect the infection dynamics \citep{dietz1988std}. For example, long-term, monogamous relationships protect the couple from infection if both are uninfected and slows the spread if one or both members of the couple are infected \citep{Morris1995Concurrent,Leng2018ConcurrencyControl,Kretzschmar1996ConcurrencyModel}. High concurrency, i.e., having more than one sexual partner at the same time, and casual sexual encounters can accelerate the spread of infections. A plethora of models have been developed to assess the dynamics of sexual contact networks and the infections that spread upon them \citep{Mattie2025NetworkModels}, including pair formation models, (temporal) statistical network models, and mechanistic or agent-based models.

Pair formation models based on ordinary differential equations (ODEs) are a natural extension of compartmental disease models to account for persistent contact patterns between individuals. They were originally developed to investigate the spread of HIV among populations of men who have sex with men (MSM), evolving from basic pair formation models \citep{Kretzschmar1996ConcurrencyModel} to include migration \citep{Kretzschmar1998PairFormation} and different contact types \citep{Xiridou2003AmsterdamSteadyCasual}. Many studies assume serial monogamy \citep{Kretzschmar1998PairFormation,Hansson2019StockholmModel} or closed populations without migration \citep{Leng2018ConcurrencyControl}.

These models are interpretable and demonstrate the importance of accounting for the structure of contact networks for investigating the spread of STIs. However, estimating model parameters from data is challenging. ODE models are inherently deterministic such that they cannot account for either noise in the processes that govern network evolution or the data collection process. Typically, point estimates of parameters are obtained by matching summary statistics of surveys and model predictions. \citet{Xiridou2003AmsterdamSteadyCasual} and \citet{Kretzschmar1998PairFormation} conducted sensitivity studies by varying model parameters, but principled inference is required to make decisions under uncertainty, such as prioritizing different interventions. Further, the coupled ODEs quickly become intractable for more complex network evolution mechanisms.

Statistical network models, especially (temporal) exponential random graph models (ERGMs), address uncertainty quantification for contact networks \citep{Jenness2018EpiModel}. In principle, the likelihood of the models is tractable and expressed in terms of the number of dyads, triangles, or other network motifs---the sufficient statistics of the ERGM. However, in practice, inference is often complicated by the difficulty of evaluating the normalizing constant of the likelihood, and sophisticated techniques have been developed to address this challenge \citep{Chatterjee2013ERGMs}. While ERGMs are interpretable in terms of network motifs, they do not naturally capture the processes that govern network evolution, making it more difficult to test interventions in silico or to consider counterfactuals in causal inference settings.

Agent-based models simulate the evolution of sexual contact networks based on simple rules that are interpretable, enable integration of domain knowledge, and facilitate in-silico assessment of interventions by modifying the simulation. However, learning the parameters of these models is challenging because the likelihood is intractable and collecting high-quality data on sexual contact networks is difficult.

These different model classes have been used to study a range of infections (including HIV \citep{Hansson2019StockholmModel,Kretzschmar1998PairFormation}, chlamydia \citep{Roenn2023Chlamydia}, gonorrhea, HPV, and mpox \citep{Xiu2025mpoxCanada,Crenshaw2025Mpox}) and populations (including unipartite networks representing homosexual partnerships \citep{Kretzschmar1996ConcurrencyModel,Kretzschmar1998PairFormation,Hansson2019StockholmModel,Xiridou2003AmsterdamSteadyCasual}, bipartite networks to investigate heterosexual partnerships \citep{Leng2018ConcurrencyControl}, and more complex, structured networks to account for covariates such as age \citep{Roenn2023Chlamydia} and race \citep{Beck2015AgeAndRace}). However, marrying interpretable simulations that facilitate assessment of interventions in silico with principled inference for parameters of network models remains challenging.

To tackle the tension between mechanistic simulation and principled inference, we introduce a framework that enables robust Bayesian inference for flexible, agent-based network models based on longitudinal network data. 
First, we develop a discrete-time simulator that specializes to many existing models for certain parameter values, and we derive the fraction of individuals in a steady relationship in the thermodynamic limit, i.e., in the limit of a very large population. 
Second, we use approximate Bayesian computation to infer the parameters of the model from summary statistics reported in a recent study of 403 MSM in Sweden \citep{Hansson2019StockholmModel}, offering principled uncertainty quantification.
Third, we construct longitudinal summary statistics based on only binary questions that are easy for respondents to answer and are less likely to suffer from recall bias than complex survey instruments, such as diaries of sexual activity. We demonstrate that these summaries can significantly enhance parameter inference---although at the cost of conducting multiple survey waves. The degree of extra information obtained from longitudinal summaries depends on the follow-up period, and simulations can inform the design of future surveys.

\section{Network Model}

We require an algorithm that can simulate realistic networks given the parameters we seek to infer. We develop a discrete-time, stochastic algorithm that evolves the network using a set of mechanisms or rules. Employing a discrete-time algorithm simplifies the interpretation and implementation compared with stochastic continuous-time models that often rely on complex stochastic differential equations. We used an interval of one week. This is fast compared to the timescale of most of the mechanisms we consider, and it allows us to approximate continuous dynamics well.

\begin{figure}
    \centering
    \includegraphics[width=\textwidth]{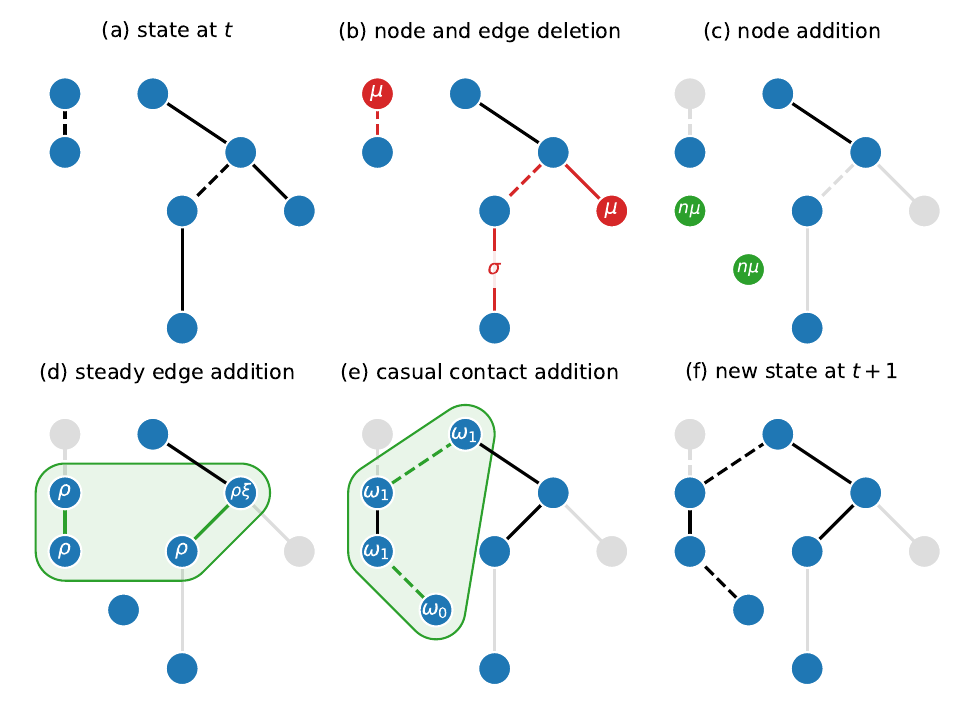}
    \caption{\emph{Simple mechanistic rules of the model give rise to complex network dynamics.} Panel~(a) shows a sexual contact network at step $t$, where nodes, solid edges, and dashed edges represent individuals, steady relationships, and casual encounters, respectively. In panel~(b), all casual encounters are removed, steady relationships are removed with probability $\sigma$, and nodes are removed with probability $\mu$. If a node is removed, all its steady relationships are also removed. New nodes are added in panel~(c), where the number of new nodes is sampled from a Poisson distribution with rate $n\mu$. As shown in panel~(d), individuals enter a pool of steady-relationship-seekers with probability $\rho\xi^k$, where $k$ is the current degree (the number of current steady relationships) of each node. Nodes in the pool are randomly paired. Similarly, as illustrated in panel ~(e), individuals enter a pool of casual-encounter-seekers with probability $\omega_0$ if they are single and $\omega_1$ if they are not. The four-step process gives rise to the new state at step $t+1$ in panel~(f).}
    \label{fig:illustration}
\end{figure}

Our model builds on and consolidates several well-understood network evolution mechanisms.The network $G=\parenth{V,E_s,E_c}$ comprises sets of nodes $V$, steady edges $E_s$, and casual edges $E_c$. Each node $i\in V$ represents an individual who can interact with others. We are interested in a unipartite network of edges between MSM, and we do not consider any node attributes, although mixing patterns by age and race may be important \citep{Beck2015AgeAndRace} and accounting for sex is essential for the study of heterosexual populations \citep{dietz1988std,Leng2018ConcurrencyControl}. A steady edge $\parenth{i,j}\in E_s$ represents a persistent, undirected relationship between two individuals $i$ and $j$ \citep{dietz1988std}. Nodes may have more than one steady edge, i.e., steady relationships may be concurrent. Following \citet{Xiridou2003AmsterdamSteadyCasual} and \citet{Hansson2019StockholmModel}, a casual edge $\parenth{i,j}\in E_c$ represents a one-time sexual encounter, although the definition of what constitutes a casual edge follows different conventions. For example, \citet{weiss2020artnet} use casual edges to represent persistent, but short-lived connections and reserve the notion of one-time edges to capture instantaneous interactions.

In our model, at each step $t$ of the simulator, the current state of the network, shown in panel~(a) of \cref{fig:illustration}, evolves in four phases. First, as shown in panel~(b), existing nodes leave the population with probability $\mu$, representing emigration, aging out of the sexually active population, and death. Any edges these nodes had are removed from the network. All casual edges are removed because they represent instantaneous interactions. Remaining steady edges are dissolved with probability $\sigma$, representing the natural breaking up of relationships.

Second, we sample the number of new nodes to be added to the network from a Poisson distribution with rate $n\mu$, representing immigration and aging into the sexually active population, as shown in panel~(c). The expected population size is thus $n$. We use integers to label nodes, and the labels of new nodes are monotonically increasing to ensure distinct identities in the simulation.

Third, singles seek a new steady relationship with probability $\rho$, i.e., they enter a pool of candidates to be paired up, as shown in panel~(d). Individuals who are not single enter the pool with probability $\rho \xi^k$, where $k$ is the degree, i.e., the number of steady edges the individual already has, and $\xi$ is a concurrency parameter. On the one hand, if $\xi=0$, a node with an existing steady edge will never seek another, i.e., the population is serially monogamous. On the other hand, if $\xi=1$, the behavior for seeking new steady edges is independent of the number of existing ones. For intermediate $0<\xi<1$, the tendency to seek new steady edges decreases with the degree, reflecting the limited capacity for nodes to have multiple steady edges. This approach closely follows \citep{Kretzschmar1996ConcurrencyModel}, although they omitted the $k$ exponent. Once all nodes have entered the pool, they are randomly paired up with each other. If the number of nodes in the pool is odd, one is dropped at random.

\begin{algorithm}
    \caption{Stochastic discrete-time model which reduces to previous proposals for specific parameter values. The function $\textsc{random\_pairs}\parenth{C}$ transforms a set of nodes $C$ into an edge set of randomly paired nodes, discarding a random node if $\abs{S}$ is odd; $\textsc{is\_single}\parenth{i}$ returns whether $i$ has no steady edges.}
    \label{alg:universal-simulator}
    \begin{algorithmic}
        \Require $G=\parenth{V,E_s,E_c}$ \Comment{Graph with vertices $V$, steady and casual edges $E_s$ and $E_c$.}
        \Require $n>0$ \Comment{Expected number of nodes.}
        \Require $0\leq \mu \leq 1$ \Comment{Probability for a node to leave the population.}
        \Require $0\leq \rho \leq 1$ \Comment{Probability for a node to seek a steady relationship.}
        \Require $0\leq \xi \leq 1$ \Comment{Concurrency parameter with $\xi=0$ indicating serial monogamy.}
        \Require $0\leq \sigma \leq 1$ \Comment{Probability for a steady relationship to dissolve.}
        \Require $0 \leq \omega_0 \leq 1$ \Comment{Probability to seek casual partner if single.}
        \Require $0 \leq \omega_1 \leq 1$ \Comment{Probability to seek casual partner if not single.}
        \Function{step}{$G,n,\mu,\rho,\xi,\sigma,\omega_0,\omega_1$}
            \State $\parenth{V,E_s,E_c}\gets G$ \Comment{Unpack vertices and edges.}
            \Statex

            \State \hrulefill{} \textbf{Remove existing nodes and create new ones.} \hrulefill
            \State $j\gets\max V$ \Comment{Store largest node index.}
            \State $V\gets\set{i\in V\mid x_i=0 \text{ for }x_i\sim \mathsf{Bernoulli}\parenth{\mu}}$ \Comment{Remove nodes.}
            \State $r\sim\mathsf{Poisson}\parenth{\mu n}$ \Comment{Sample number of new nodes.}
            \State $V\gets V\cup\set{j + 1, \ldots, j + r}$ \Comment{Add new nodes.}
            \Statex

            \State \hrulefill{} \textbf{Dissolve steady edges and create new ones.} \hrulefill
            \State $E_s\gets \set{e \in E_s\mid x_e=0\text{ for }x_e\sim\mathsf{Bernoulli}\parenth{\sigma}}$ \Comment{Dissolve steady edges.}
            \State $S\gets \set{i\in V\mid x_i=1 \text{ for }x_i\sim\mathsf{Bernoulli}\parenth{\rho \xi^k}}$ \Comment{Select candidates for new steady edges modulated by concurrency parameter $\xi$, where $k$ is the degree of steady edges.}
            \State $E_s\gets E_s\cup\Call{random\_pairs}{S}$ \Comment{Create new steady edges.}
            \Statex

            \State \hrulefill{} \textbf{Dissolve casual edges and create new ones.} \hrulefill
            \State $C\gets \set{i\in V\mid x_i=1\text{ for }x_i\sim\mathsf{Bernoulli}\parenth{\omega_0\text{ if }\Call{is\_single}{i}\text{ else }\omega_1}}$ \Comment{Select candidates for new casual edges.}
            \State $E_c\gets \Call{random\_pairs}{C}$ \Comment{Replace casual edges with new ones.}
            \Statex

            \State $G\gets\parenth{V,E_s,E_c}$ \Comment{Package vertices and edges.}
            \State \Return $G$
        \EndFunction
    \end{algorithmic}
\end{algorithm}

Fourth, nodes seek casual edges with probability $\omega_0$ if they are single (not in a steady relationship) and probability $\omega_1$ if they are not \citep{Hansson2019StockholmModel,Xiridou2003AmsterdamSteadyCasual}. Nodes are again randomly paired after establishing the pool of candidates, as shown in panel~(e). We arrive at the new state shown in panel~(f) after applying the rules controlled by seven parameters: expected population size $n$, probability for a node to leave the population $\mu$, probability to seek a steady relationship $\rho$, probability for a steady relationship to dissolve naturally $\sigma$, concurrency parameter $\xi$, probability to seek casual contacts for singles $\omega_0$, and probability to seek casual contacts for partnered individuals $\omega_1$. The simulation allows for both concurrency of steady edges \citep{Kretzschmar1996ConcurrencyModel} and casual contacts \citep{Hansson2019StockholmModel,Xiridou2003AmsterdamSteadyCasual}. Pseudocode for the model is shown in \cref{alg:universal-simulator}. Like all stochastic models, the network evolution has a transient burn-in period before reaching steady state. To mitigate this issue, we run all simulations for 30~years at weekly intervals (1{,}560 steps) before subsequent analysis. 

\subsection{Properties in the Thermodynamic Limit}\label{sec:analytics}

A key quantity of pair-formation models is the fraction of individuals $f_t$ who are in a steady relationship at time $t$. 
While this quantity is difficult to derive in closed form for the general model, we can make progress by making two assumptions. First, we consider the thermodynamic limit, i.e., the limit of an infinite population size. Second, we assume serial monogamy with $\xi=0$ such that each node has at most one steady partner.
As shown in \cref{alg:universal-simulator}, the simulator first accounts for relationship dissolution and subsequently creates new relationships. In contrast to continuous-time models, this ordering is significant for discrete-time simulators.
At each step, a steady relationship survives with probability $\parenth{1-\mu}^2\parenth{1-\sigma}$, where the first term accounts for neither of the two partners leaving the population and the latter accounts for the relationship not naturally dissolving. Based on a geometric distribution, the expected relationship length is thus
\begin{equation}
    \frac{1-\parenth{1-\mu}^2\parenth{1-\sigma}}{\parenth{1-\mu}^2\parenth{1-\sigma}}\label{eq:relationship-length}.
\end{equation}
After the dissolution step, the remaining fraction of partnered individuals is thus $f_t' = f_t \parenth{1-\mu}^2\parenth{1-\sigma}$, and the fraction of singles is $1-f'_t$. Each of the singles seeks out a relationship with probability $\rho$, and the fraction of partnered individuals at the next step is
\begin{align}
    f_{t+1} &= f'_t + \rho\parenth{1-f'_t}\nonumber\\
    &=\parenth{1-\mu}^2\parenth{1-\rho}\parenth{1-\sigma}f_t + \rho.\label{eq:frac-paired-evolution}
\end{align}
In steady state, $f_{t+1}=f_t$, and solving \cref{eq:frac-paired-evolution} yields the steady-state fraction of paired individuals
\begin{equation}
f=\frac{\rho}{1-\alpha}\text{, where }\alpha=\parenth{1-\mu}^2\parenth{1-\rho}\parenth{1-\sigma}.\label{eq:steady-state-f}
\end{equation}

However, if we follow a cohort of individuals and interview them again $\tau$ steps later, the expected fraction $g_\tau$ of paired individuals in the \emph{remaining} cohort changes over time. By construction, members of the remaining cohort have not emigrated so their relationships remain intact with probability $\parenth{1-\mu}\parenth{1-\sigma}$, where the two factors account for their partner leaving and the relationship dissolving naturally, respectively. Following the same steps as above, we find
\[
g_{\tau+1}=\beta g_\tau + \rho\text{, where } \beta=\parenth{1-\mu}\parenth{1-\sigma}\parenth{1-\rho}.
\]
We can solve the recurrence relation subject to the initial condition $g_0=f$ to obtain
\begin{equation}
    g_{\tau} = \frac{\rho}{1-\beta}\bracket{1 - \beta^{\tau + 1}\parenth{\frac{\mu}{1-\alpha}}}.\label{eq:steady-state-g}
\end{equation}
For any lag $\tau>1$ and non-zero emigration probability $\mu$, the fraction of paired individuals $g_\tau$ in the remaining cohort is larger than the fraction of paired individuals $f$ in the general population because the probability for a relationship with a member of the cohort to survive is larger by a factor of $1-\mu$. At long lags, $\lim_{\tau\rightarrow\infty}g_\tau=\frac{\rho}{1-\beta}$ because $\beta<1$ such that the second term in brackets vanishes.

\begin{figure}
    \centering
    \includegraphics[width=\textwidth]{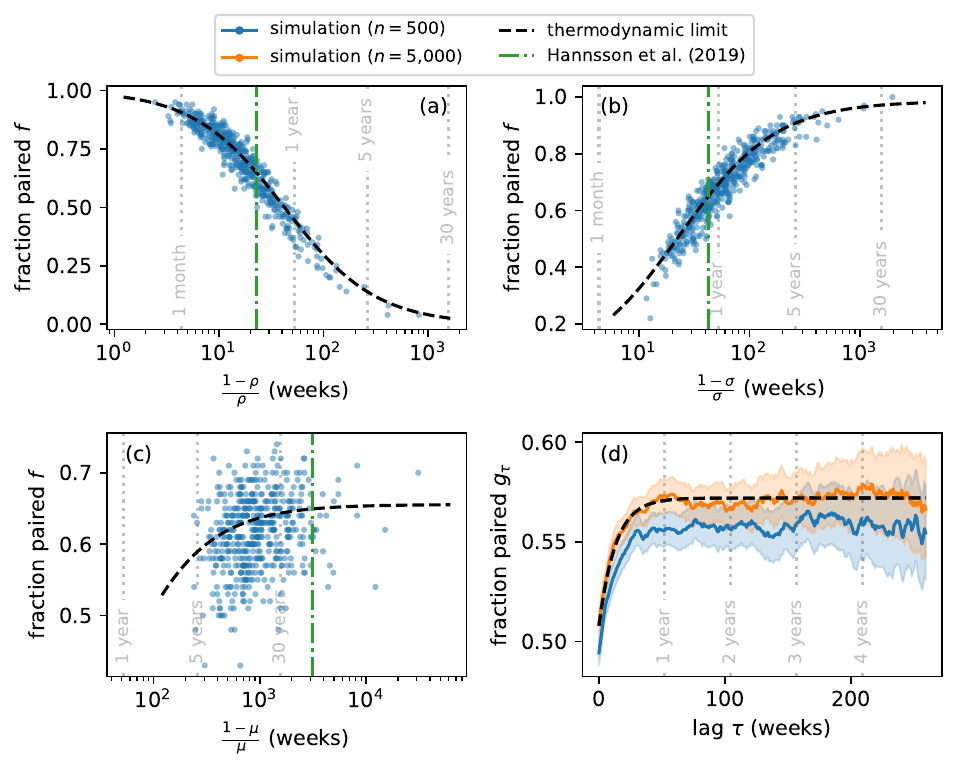}
    \caption{
    \emph{Analytic results in the thermodynamic limit approximate finite-size simulations well.} 
    Panels~(a--c) show the dependence of the fraction of paired nodes $f$ on relationship formation parameter $\rho$, dissolution parameter $\sigma$, and migration parameter $\mu$, respectively. For each panel, the other parameters are fixed at values reported by \citet{Hansson2019StockholmModel} (dot-dashed vertical line for the parameter under consideration). Simulations with expected population size $n=500$ and $m=100$ survey respondents (points) are in close agreement with analytic expressions in the thermodynamic limit (dashed lines) for $\rho$ and $\sigma$. There is no clear relationship for $\mu$ because it only has a small effect on $f$ over the plausible parameter range. 
    For interpretability, model parameters have been transformed from weekly probabilities to the expected time for an event to occur.
    Panel~(d) shows the fraction of paired nodes in the cohort $g_\tau$ as a function of lag $\tau$ between survey waves. The two lines and shaded bands represent the mean and standard error of 500 independent simulations for $n=500$ and $n=5{,}000$. Parameters are fixed at values reported by \citet{Hansson2019StockholmModel} except an elevated migration probability $\mu=0.01$ was used to illustrate the evolution of $g_\tau$. There is a bias for small $n$ (see main text for details), and the standard error increases with $\tau$ due to emigration.
    }
    \label{fig:thermodynamic}
\end{figure}

To assess how well the fraction of paired nodes derived in the thermodynamic limit approximates finite-size simulations, we ran simulations with expected population size $n=500$ and a cohort size of $m=100$ respondents. For each of $\rho$, $\sigma$, and $\mu$, we sampled parameter values from a plausible range informed by previous studies (see \cref{sec:prior-elicitation} for details), held the other parameters fixed at the values reported by \citet{Hansson2019StockholmModel}, and ran 500 independent simulations. The dependence of the fraction of paired nodes $f$ on the three parameters is shown in panels (a--c) of \cref{fig:thermodynamic}. The relationships demonstrate close agreement between simulations and the analytic expression in \cref{eq:steady-state-f} for $\rho$ and $\sigma$---despite the small expected population size. The relationship is less apparent for the migration parameter $\mu$ because the fraction of paired individuals does not vary appreciably over the domain of plausible values. 

Panel~(d) shows the mean and standard error of the fraction of paired nodes in the cohort for 500 simulations as a function of the follow-up lag $\tau$. Parameters were fixed at values reported by \citet{Hansson2019StockholmModel}, except we used an elevated migration probability $\mu=0.01$ for illustration.
While the analytic expression follows the same trend as the simulations, the former are overestimates of the latter. This is a consequence of the cohort being a non-negligible fraction of the population. Under the assumption of an infinite population used to derive \cref{eq:steady-state-g}, the probability for a steady relationship to be between two members of the cohort is vanishingly small. But, in our simulations, the initial cohort makes up 20\% of the population such that removing a node from the cohort is likely to lead to loss of a steady partnership for another cohort member, resulting in a reduced fraction of partnered individuals in finite-size simulations. Increasing the expected population size to $n=5{,}000$ attenuates the bias. We also observe that the standard error increases with $\tau$ because the size of the cohort decreases with time, giving rise to larger variability.

\section{Parameter Inference}

The likelihood of the model is not tractable, and we have to resort to simulation-based inference. We employ approximate Bayesian computation (ABC) because summary statistics of sexual contact networks are readily collected in surveys about sexual behavior \citep{weiss2020artnet,Hansson2019StockholmModel,Xiridou2003AmsterdamSteadyCasual}. ABC samples from an approximate posterior distribution by comparing observed data with simulations in three steps \citep{Beaumont2002RegressionAdjustment}. First, we sample parameters from the prior distribution $p\parenth{\theta}$ and run the simulator $p\parenth{G\mid\theta}$ for each sample to obtain a large number of samples $\parenth{\theta_i,G_i}$ from the prior predictive distribution. We use $\theta$ to collectively denote the set of all parameters for conciseness. Second, we evaluate the distance $d_i=d\parenth{G_i, G_*}$ between the observed graph $G_*$ and the $i^\text{th}$ simulated graph $G_i$. Finally, we accept $\theta_i$ as a sample from the approximate posterior $\bar p\parenth{\theta\mid G_*}$ if the distance $d_i$ is smaller than a threshold $\epsilon$. The smaller $\epsilon$, the better the approximation. Intuitively, ABC draws parameter samples $\theta_i$ that
generate simulated graphs $G_i$ which ``look like'' the observed graph $G_*$. In practice, we do not observe the full graph $G_*$, and defining a distance measure $d$ is non-trivial even if we did \citep{Hoffmann2025summaries}. We thus evaluate summary statistics $t_i=t\parenth{G_i}$ of the simulated graphs and compare them with data collected in network surveys $t_*$, e.g., the fraction of nodes with at least one steady edge, the average length of relationships, or the frequency of casual contacts.
Neural posterior density estimation is a data-efficient alternative to ABC \citep{Papamakarios2016epsilonFree}, and graph neural networks are effective architectures for inference \citep{Hoffmann2025summaries}. However, graph neural networks require access to the graph structure, not just summary statistics collected in surveys, and we do not consider them here.

\begin{figure}
    \centering
    \includegraphics[width=\textwidth]{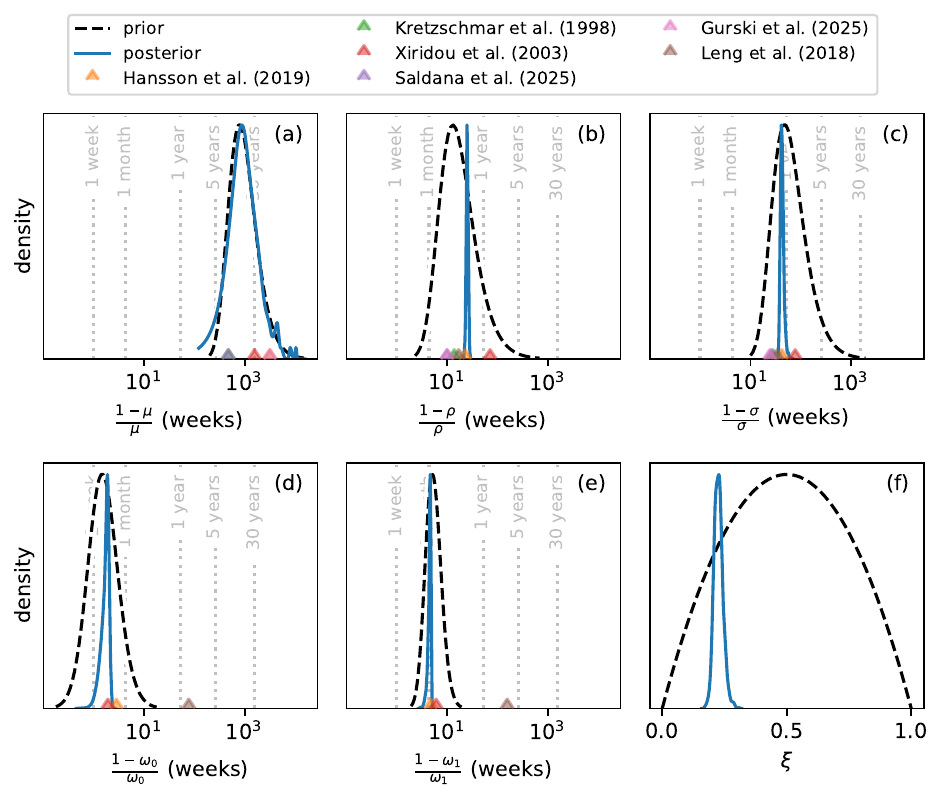}
    \caption{\emph{Approximate Bayesian computation yields high-fidelity posteriors based on survey data.} The dashed and solid lines in each panel represent the prior and posterior after fitting to data from \citet{Hansson2019StockholmModel}, respectively. Markers in the rug plot show point estimates of parameter values from the literature. The posterior is consistent with point estimates from \citet{Hansson2019StockholmModel} but offers principled uncertainty quantification. The model is specified in terms of event probabilities per week. However, for interpretability in panels~(a--e), we visualize the distributions of the transformed parameter expressed in terms of inverse odds which corresponds to the expected time for an event to occur. The concurrency parameter $\xi$ has no temporal interpretation and is shown on its natural scale.}
    \label{fig:parameters}
\end{figure}

\subsection{Prior Elicitation}\label{sec:prior-elicitation}

Since we take a Bayesian perspective which requires the definition of priors over all parameters. We elicited priors based on previous literature, as shown in \cref{fig:parameters}. We use beta priors for all model parameters except the expected population size $n$. However, for interpretability, we present beta-prime priors on $\frac{1-\lambda}{\lambda}$, where $\lambda$ denotes one of the probabilities. This is the expected number of steps until the next event. The hyperparameters for beta distributions were chosen empirically so none of the parameter values reported in the literature, shown as markers of a rug plot in each panel of \cref{fig:parameters}, are too far in the tail of the prior. For models that reported parameters as rates of continuous-time ODE models and as probabilities of discrete-time models not on a weekly timescale, we converted parameters to the probability of an event occurring per week, as outlined in \cref{app:conversion}. 

The probabilities $\omega_0$ and $\omega_1$ for forming casual contacts each week that were reported by \citet{Leng2018ConcurrencyControl} are an exception: The probabilities are unexpectedly small compared with other studies which is likely explained by a different study population and methodology. First, in contrast to other studies focused on MSM, \citet{Leng2018ConcurrencyControl} consider a heterosexual population aged between 16 and 25 years of age. Second, they ``simply [captured] whether or not individual have been involved in a concurrent partnership'' to estimate the rate of casual contacts, i.e., they used concurrency of \emph{steady} edges as a proxy for \emph{casual} contacts.

Previous literature does not report values for the concurrency parameter $\xi$, although \citet{Kretzschmar1996ConcurrencyModel} conducted a sensitivity analysis by varying $\xi$ and assessing the impact on network statistics. We use a relatively uninformative $\mathsf{Beta}\parenth{2,2}$ prior, reflecting that neither all people are monogamous nor are all people pursuing an arbitrary number of steady relationships.

\subsection{Summary Statistics}

To generate approximate posterior samples using ABC, we need to collect summary statistics reported in surveys and evaluate these same summary statistics for each of the simulated graphs $G_i$. We base our analysis on a survey of 403 MSM in Stockholm, Sweden conducted in 2015 \citep{Hansson2019StockholmModel}. They report the fraction of respondents with at least one steady relationship (64\%), the fraction of respondents who report concurrent steady relationships (14.6\%), the average length of steady relationships (203~days), the average duration between consecutive casual contacts for singles (23.1~days) and partnered individuals (36.3~days). Data were collected using a timeline follow-back setup over a 12~month period, i.e., respondents were asked to recall sexual relationships with different partners over a one-year time frame. Consequently, duration estimates are right-censored at one year.
\citet{weiss2020artnet} also report comprehensive summary statistics for almost five thousand MSM in the United States, but the publicly available data do not include information on the rate of casual encounters broken down by relationship status.

\begin{figure}
    \centering
    \includegraphics[width=\textwidth]{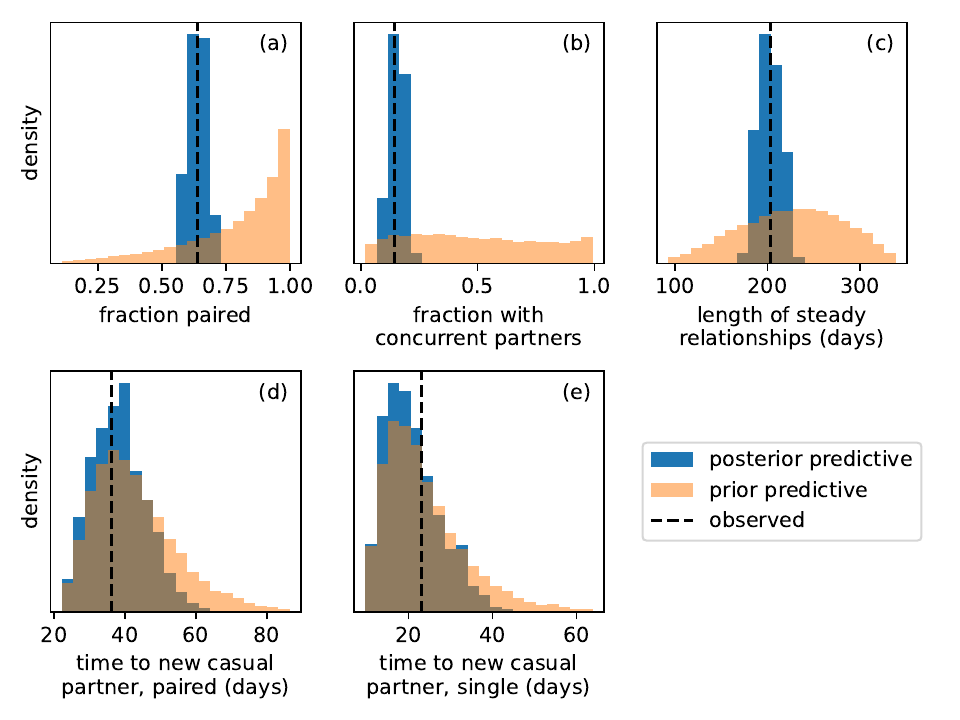}
    \caption{\emph{Replicated samples from the posterior predictive distribution are consistent with the observed data.} Panels~(a--e) show the data observed by \citet{Hansson2019StockholmModel} as a dashed, vertical line, the prior predictive distribution of each summary as a transparent histogram, and the posterior-predictive replication of summaries after fitting as a solid histogram.}
    \label{fig:posterior-predictive-replication}
\end{figure}

As shown in \cref{sec:analytics}, the expected value of several summary statistics does not depend on the expected population size $n$, but the sample size affects the sampling distribution of summary statistics. To mitigate this effect, we ran all simulations with $n=5{,}000$, an expected population size more than an order of magnitude larger than the sample size of the survey and comparable with the MSM population in Stockholm. \citet{Marcus2013MsmPopulationEstimates} estimated the population of MSM in Sweden as approximately 65{,}000, and approximately 10\% of the Swedish population live in Stockholm, although the density of MSM may be larger in Stockholm than the rest of the country. Samples from the prior predictive distribution of summary statistics together with the observed summary statistics are shown in \cref{fig:posterior-predictive-replication}.

\subsection{Approximate Bayesian Computation and Regression Adjustment}

We generated $10^5$ synthetic graphs to build a reference table of summary statistics. To bring all summary statistics onto the same scale, we divided all durations by the number of days per year such that all summaries are on the unit interval. We evaluated the Euclidean distance between summary statistics and accepted 1\% of the reference table with the smallest distance as samples from the approximate posterior distribution $\bar p\parenth{\theta\mid t_*}$.

\newcommand{\ci}[2]{(#1--#2)}
\begin{table}
    \newcolumntype{.}{D{.}{.}{2}}
    \centering
    \begin{tabular}{l.@{ }c..}
        \toprule
        &\multicolumn{4}{c}{\textbf{Posterior Quantiles}}\\
        \cmidrule(lr){2-5}
        \textbf{Parameter}&\multicolumn{2}{c}{Median}&\colalign{c}{2.5\textsuperscript{th}}&\colalign{c}{97.5\textsuperscript{th}}\\
        \midrule
        Migration $\frac{1-\mu}{\mu}$&15.7&years &5.9&80.5\\
        Steady edge formation $\frac{1-\rho}{\rho}$&24.9&weeks&22.8&27.3\\
        Steady edge dissolution $\frac{1-\sigma}{\sigma}$&42.2&weeks&37.9&51.4\\
        Casual edge formation (singles) $\frac{1-\omega_0}{\omega_0}$&1.8&weeks&1.0&2.1\\
        Casual edge formation (partnered) $\frac{1-\omega_1}{\omega_1}$&4.5&weeks&3.6&4.8\\
        Concurrency $\xi$&0.23&---&0.19&0.27\\
        \bottomrule
    \end{tabular}
    \caption{Posterior median and 95\% credible interval for parameters of the mechanistic network model.}
    \label{tbl:estimates}
\end{table}

To further improve the posterior samples, we applied regression adjustment \citep{Beaumont2002RegressionAdjustment}. Simulated summaries $t_i$ and observed summaries $t_*$ never match exactly, and regression adjustment ``corrects'' the parameter samples $\theta_i$ for this mismatch. In short, a multivariate linear regression model $f$ is fit to predict the parameters $\theta$ given summaries $t$ based on the selected samples. The parameter samples are then adjusted by $\theta_i \rightarrow \theta_i + f\parenth{t_*} - f\parenth{t_i}$ as if the simulated summaries had matched exactly, compensating for the mismatch. Regression adjusted posterior samples are shown as a kernel density estimate in \cref{fig:parameters} and quantiles are listed in \cref{tbl:estimates}. The posteriors are consistent with the point estimates reported by \citet{Hansson2019StockholmModel} but provide principled uncertainty estimates.

The typical timescale for individuals seeking a new steady relationship is 24.9~weeks with 95\% credible interval (CI) 22.8--27.3~weeks, and the natural timescale for steady edges dissolving is 42.2~weeks (95\% CI 37.9--51.4 weeks). The observed relationship lengths are shorter because steady edges also end if at least one partner leaves the population, as indicated by \cref{eq:relationship-length}. The timescale for having casual contacts is faster for singles at 1.8~weeks (95\% CI 1.0--2.1 weeks) than individuals with steady edges at 4.5~weeks (95\% CI 3.6--4.8). Individuals who already have one steady edge are only $\xi=23\%$ as likely as singles to seek a new steady edge (95\% CI 19--27\%), reflecting the tendency for individuals to only have one steady partner at a time.

The summary statistics contain no information about the migration parameter $\mu$, and the posterior is indistinguishable from the prior in panel~(a) of \cref{fig:parameters}. Due to the lack of information, many studies assume a fixed value for the migration time scale, such as 10~years \citep{Kretzschmar1998PairFormation,Saldana2025hpv}, 30~years \citep{Xiridou2003AmsterdamSteadyCasual}, or 60~years \citep{Hansson2019StockholmModel,Gurski2025pairformation}. Others consider only a closed population without migration \citep{Leng2018ConcurrencyControl,Kretzschmar1996ConcurrencyModel}. As demonstrated in the following section, longitudinal summary statistics from surveys with two or more waves can identify the migration parameter $\mu$.

To assess the fit of the model and the quality of approximate posterior samples, we compared the observed data and samples from the posterior predictive distribution. As shown in \cref{fig:posterior-predictive-replication}, samples drawn from the posterior predictive distribution are consistent with the observed summaries.

\subsection{Longitudinal Summary Statistics\label{sec:longitudinal}}

Inferring the migration parameter $\mu$ is fundamentally difficult given only the data that is typically available in cross-sectional surveys. Further, eliciting reliable information through questionnaires can be difficult in light of the often complex nature of intimate relationships. For example, \citet{Hansson2019StockholmModel} capture important temporal information using a timeline follow-back approach, asking respondents to recall sexual contacts and relationships over the preceding 12~month period. Responses may suffer from recall bias, especially for individuals with a large number of contacts, e.g., up to 250~yearly partners in their study. \citet{weiss2020artnet} collect the total number of partners over the past year, but only obtain detailed information for the most recent five partnerships. In contrast, longitudinal studies, such as the Amsterdam Cohort Study \citep{Davidovich2001acs}, can elicit temporal information by asking simple questions that are easier for respondents to answer. By tracking the survey sample, longitudinal studies can also constrain the migration parameter $\mu$.

\citet{Smiley2024longitudinal} demonstrated that combining data from multiple survey waves can improve parameter inference using ABC because the effective sample size increases and the variance of summary statistics decreases. Here, we demonstrate that longitudinal summaries more naturally inform the parameters of the network evolution model. By longitudinal summaries, we mean summary statistics that compare the state of the contact network at different time points and require that nodes are identifiable across different waves of the survey. The questions to elicit longitudinal information are often simple, improving data quality.

We propose collecting two additional cross-sectional summary statistics and two longitudinal summary statistics, all based on binary questions that are easy for respondents to answer. First question: ``In the last week, did you have a casual sexual contact?'' The time span can be adjusted to better inform parameters for the target population, e.g., a longer time span for populations with a low rate of casual contacts. We use a weekly timescale because it matches our simulation, and the typical time span between casual sexual contacts among MSM is on the order of weeks. The resulting cross-sectional summary statistics, the fraction of singles and paired nodes with a casual contact, inform the weekly contact probabilities $\omega_0$ and $\omega_1$, respectively. Second, for individuals with at least one steady edge at the previous wave, we ask the following question:``Do you still have a steady relationship with [name]?'' The resulting summary statistics, the fraction of retained steady edges, is highly informative of the edge dissolution probability $\sigma$ and weakly constrains the migration parameter $\mu$. Third, we track the fraction of nodes who have left the population which directly informs the migration parameter $\mu$, although this statistic may be confounded by survey drop-out rather than true migration. Because emigration and natural relationship dissolution are independent events across all individuals, the expected values for a follow-up lag $\tau$ are 
\begin{align}
    \mathbb{E}\bracket{\text{fraction of retained nodes}}&=\parenth{1-\mu}^\tau\label{eq:frac-retained-nodes}\\
    \mathbb{E}\bracket{\text{fraction of retained edges}}&=\bracket{\parenth{1-\mu}^2\parenth{1-\sigma}}^\tau\label{eq:frac-retained-edges}.
\end{align}
\Cref{tbl:summaries} provides an overview of all summary statistics.

\begin{table}
    \centering
    \newcolumntype{.}{D{.}{.}{2}}
    \begin{tabular}{lll.@{ }c}
        \toprule
        \textbf{Summary} & \textbf{Type} & \textbf{Expectation} & \multicolumn{2}{l}{\textbf{Observed}}\\
        \midrule
        \multicolumn{4}{c}{\emph{Cross-Sectional Summaries}}\\
        Fraction paired & binary & \cref{eq:steady-state-f} &0.64\\
        Fraction with concurrent steady edges & binary & intractable &0.15\\
        Duration of steady edges & TLFB & \cref{eq:relationship-length} & 203.2&days\\
        Gap between casual partners (single) & TLFB & intractable &23.1&days\\
        Gap between casual partners (paired) & TLFB & intractable &36.3&days\\
        Fraction singles with casual contact & binary & $\omega_0$ & \colalign{c}{---}\\
        Fraction paired with casual contact & binary & $\omega_1$ & \colalign{c}{---}\\
        \midrule
        \multicolumn{4}{c}{\emph{Longitudinal Summaries}}\\
        Fraction of retained nodes & binary & \cref{eq:frac-retained-nodes} &\colalign{c}{---}\\
        Fraction of retained edges & binary & \cref{eq:frac-retained-edges} &\colalign{c}{---}\\
        \bottomrule
    \end{tabular}
    \caption{Overview of cross-sectional and longitudinal summary statistics. Observed values are from \citet{Hansson2019StockholmModel}. Binary questions are simple yes-no questions, and timeline follow-back (TLFB) questions ask respondents to complete a timeline of all relationships over a fixed period, typically 6~to 12~months. An analytic expression for the expected fraction of paired nodes is only tractable in the thermodynamic limit for serial monogamy, i.e., $\xi=0$. The expected duration length is only tractable if relationship lengths are not right-censored as in TLFB questionnaires.}
    \label{tbl:summaries}
\end{table}

\begin{figure}
    \centering
    \includegraphics[width=\textwidth]{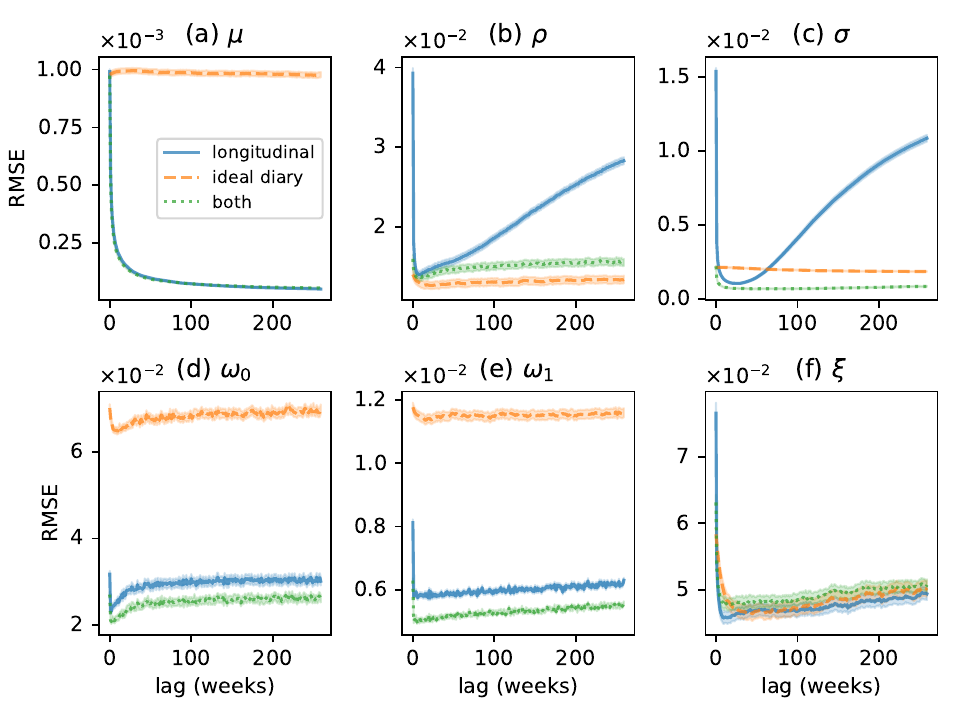}
    \caption{\emph{The quality of posterior inference depends on the lag between subsequent survey waves.} Each panel shows the root-mean squared error (RMSE) of posterior samples compared against a test set of 1{,}000 simulations with known parameters sampled from the prior predictive distribution. Error bands represent the standard error across the test set. Solid, dashed, and dotted represent results using only longitudinal summary statistics, summaries based on an ideal diary of interactions as collected by \citet{Hansson2019StockholmModel}, and both sets of summaries, respectively. Panels~(a--e) represent the probability to leave the population, seek a new steady relationship, for a steady relationship to dissolve, to seek a casual encounter as a single, and the probability to seek a casual encounter as a partnered individual. Panel~(f) represents the parameter controlling concurrency of steady relationships. See the main text for a detailed description and interpretation.}
    \label{fig:longitudinal-rmse}
\end{figure}

To demonstrate the utility of these longitudinal summary statistics, we created a test set of 1{,}000 simulated networks by sampling from the prior predictive distribution and inferred the parameters using ABC for varying lags between consecutive survey waves. For all cross-sectional summaries, such as the fraction of paired nodes, we took the average across survey waves, weighted by the number of respondents. The number of respondents may decrease due to migration and study drop-out. \Cref{fig:longitudinal-rmse} shows the root-mean squared error (RMSE) as a function of lag for each parameter in a separate panel. Each panel includes RMSEs inferred based on only binary questions with longitudinal summaries, an ideal diary of sexual contacts as collected by \citet{Hansson2019StockholmModel}, and all summaries. The diary is idealized because it assumes perfect recall, even for complex TLFB questionnaires.

As before, the migration parameter $\mu$ is not constrained by summaries collected through TLFB. The RMSE remains constant independent of lag, as shown in panel~(a) of \cref{fig:longitudinal-rmse}. In contrast, longitudinal summaries, specifically the fraction of retained nodes, is highly informative of $\mu$. The RMSE drops rapidly with increasing lag because the summary becomes more informative as a non-negligible fraction of the population has migrated. Using all summaries offers no further reduction in RMSE.

While confounded by $\mu$, the steady relationship length is very informative of the relationship dissolution parameter $\sigma$, and the RMSE is low for TLFB summaries independent of lag. Similar to $\mu$, the RMSE of $\sigma$ decreases rapidly with increasing lag as the fraction of retained edges becomes more informative, as shown in panel~(c). However, the characteristic timescale of natural relationship dissolution is approximately 40~weeks, as shown in \cref{tbl:estimates}. Hence, as the lag continues increasing, the summary becomes less informative because a majority of steady edges have dissolved, and the RMSE starts to increase again. We expect to see the same pattern for the RMSE of $\mu$ once the lag increases beyond 15~years, a timescale beyond the scope of our study.

The fraction of paired nodes informs the relationship formation parameter $\rho$ but also depends strongly on $\sigma$ and weakly on $\mu$ because migration occurs on a much longer timescale than natural relationship dissolution. Hence, as shown in panel~(b), the RMSE for $\rho$ at first decreases with increasing lag as constraining $\sigma$ breaks the degeneracy. TLFB immediately constrains $\sigma$ and hence offers good inference even with only a single wave. Combining all nine summaries yields inferior inferences: Adding more summary statistics increases the dimensionality, making it harder to find simulations $t_i$ that closely match the observation $t_*$ with very small $d\parenth{t_i,t_*}$.

Despite the simplicity of asking for the presence of a casual sexual contact in the preceeding week, these summaries are more informative of the casual contact parameters $\omega_0$ and $\omega_1$ than the more complex summary of the gap between contacts collected in TLFB questionnaires, as shown in panels~(d) and~(e). In both settings, the RMSE drops sharply for non-zero lag because, under the model, the probability to have casual contacts are independent across weeks, i.e., the sample size increases when more than one survey wave is conducted. However, as the lag continues increasing, respondents migrate out of the population leading to a smaller sample size and an increasing RMSE for both parameters. The increase is faster for the parameter $\omega_0$ for single than $\omega_1$ for partnered individuals---a consequence of the fraction of partnered individuals in the sample increasing, as discussed in \cref{sec:analytics}. Combining all summaries leads to a slight improvement beyond using either set of summaries independently.

The RMSE of the concurrency parameter $\xi$ decreases in line with improved inference of the other parameters that govern the formation and dissolution of steady relationships $\rho$, $\sigma$, and, to a lesser extent, $\mu$. As the ability to constrain these parameters diminishes, the RMSE of $\xi$ also increases, although the fraction of individuals with concurrent partnerships remains informative.

\section{Discussion}

We developed a discrete-time mechanistic model for sexual contact networks and an inference framework using approximate Bayesian computation. Analytic expressions for key summary statistics are derived in the steady-state limit. Parameter inference using cross-sectional survey data from 403 MSM in Stockholm yields principled uncertainty quantification for network model parameters. Further, simple longitudinal summary statistics, if available, could significantly improve inference quality for parameters that remain poorly identified by cross-sectional surveys alone.

Parameter estimates from the Stockholm survey data are consistent with point estimates reported in previous studies but provide uncertainty quantification previously unavailable. The posterior distributions reveal that some parameters are well-constrained by cross-sectional data while others remain poorly identified. Relationship formation and dissolution parameters ($\rho$ and $\sigma$) are informed by the fraction of paired individuals and relationship duration data. Casual contact parameters ($\omega_0$ and $\omega_1$) are constrained by the reported gaps between encounters for singles and partnered individuals. However, the migration parameter $\mu$ remains unidentified, with the posterior indistinguishable from the prior.

Longitudinal summary statistics based on binary questions address the parameter identifiability limitations of cross-sectional surveys. Tracking node retention between waves directly informs the migration parameter $\mu$, with inference quality improving as the lag increases and a non-negligible fraction of the population migrates. The fraction of retained relationships provides strong constraints on the dissolution parameter $\sigma$, with optimal inference occurring when the lag matches the characteristic relationship timescale. Simple weekly casual contact questions outperform complex timeline follow-back approaches for estimating contact rates.

Several limitations constrain the scope of our analysis. The analytic expressions for summary statistics assume serial monogamy ($\xi = 0$), although the simulation framework accommodates arbitrary concurrency levels and concurrency is evident in the data. While the weekly time step only approximates real-world continuous dynamics, typical relationship formation and dissolution occur on timescales of weeks to months, making this discretization appropriate. Survey dropout may confound estimates of the migration parameter $\mu$.

Future extensions could integrate infection dynamics on top of the structural network model, allowing assessment of disease spread and intervention strategies. The framework could inform optimal study design by quantifying the information gained per survey wave. Extensions to structured populations, incorporating age and race mixing patterns, would enhance realism for heterogeneous populations.
Methods for consolidating information across multiple studies with different sample sizes and summary statistics could broaden the evidence base for parameter estimation \citep{Goyal2023IntegratingNetworkData}. This framework provides a foundation for principled uncertainty quantification in network epidemiology for mechanistic network models while demonstrating the value of simple longitudinal data collection strategies.

\bibliographystyle{abbrvnat}
\bibliography{main}

\appendix

\section{Conversion of Parameter Values}\label{app:conversion}

For discrete-time models \citep{Kretzschmar1996ConcurrencyModel}, simulations may be run at different scales such as daily, weekly, or monthly scales. If an event occurs with probability $q$ on a daily basis, the probability $p$ that said event occurs at least once on a weekly basis is $p = 1 - \parenth{1 - q}^k$, where $k=7$ is the number of days per week. This expression allows us to translate between different scales. Equivalently, to translate from weekly to daily probabilities, the expression is $q = 1 - \parenth{1 - p}^{1/k}$. As an example, \citet{Kretzschmar1996ConcurrencyModel} report a daily probability $q = 0.01$ to seek a new relationship, corresponding a weekly probability $p = 0.068$. If the expected number of occurrences $q k$ is close to or exceeds one, the discrete-time simulation at the slower scale (weekly) cannot accurately approximate the dynamics at the faster scale (daily) because events are likely to occur multiple times.

For continuous-time models \citep{Kretzschmar1998PairFormation,Xiridou2003AmsterdamSteadyCasual,Leng2018ConcurrencyControl,Hansson2019StockholmModel}, parameters are typically expressed in terms of the expected number of events $x$ over a period $\tau$. Matching the first moment for a discrete-time simulation run on a time scale $\kappa$, the probability is
$$
p = \min\parenth{1, \frac{x\kappa}{\tau}},
$$
where the $\min$ ensures probabilities do not exceed one. For example, \citet{Kretzschmar1998PairFormation} report a rate of $0.73$ new partnerships per year, and the weekly probability to seek a new relationship is $p = 0.014$. If $p$ is close to one, discrete-time simulations should be run on a shorter time scale to more accurately approximate the continuous-time dynamics.

\end{document}